\documentclass[12pt]{iopart}
\usepackage{graphicx}
\usepackage{iopams}
\usepackage{url}

\begin{document}

\title{Tunable Charge Detectors for Semiconductor Quantum Circuits}

\author{C. R{\"{o}}ssler}
\address{Solid State Physics Laboratory, ETH Zurich, 8093
Zurich, Switzerland} \ead{roessler@phys.ethz.ch}
\author{T. Kr{\"{a}}henmann}
\address{Solid State Physics Laboratory, ETH Zurich, 8093
Zurich, Switzerland}
\author{S. Baer}
\address{Solid State Physics Laboratory, ETH Zurich, 8093
Zurich, Switzerland}
\author{T. Ihn}
\address{Solid State Physics Laboratory, ETH Zurich, 8093
Zurich, Switzerland}
\author{K. Ensslin}
\address{Solid State Physics Laboratory, ETH Zurich, 8093
Zurich, Switzerland}
\author{C. Reichl}
\address{Solid State Physics Laboratory, ETH Zurich, 8093
Zurich, Switzerland}
\author{W. Wegscheider}
\address{Solid State Physics Laboratory, ETH Zurich, 8093
Zurich, Switzerland}

\begin{abstract}
Nanostructures defined in high-mobility two-dimensional electron systems offer a unique way of controlling the microscopic details of the investigated device. Quantum point contacts play a key role in these investigations, since they are not only a research topic themselves, but turn out to serve as convenient and powerful detectors for their electrostatic environment. We investigate how the sensitivity of charge detectors can be further improved by reducing screening, increasing the capacitive coupling between charge and detector and by tuning the quantum point contacts' confinement potential into the shape of a localized state. We demonstrate the benefits of utilizing a localized state by performing fast and well-resolved charge detection of a large quantum dot in the quantum Hall regime.
\end{abstract}

\maketitle

\section{Introduction}
Clean low-dimensional electron systems exhibit a rich spectrum of interaction-induced effects like the formation of composite fermions~\cite{tsui_two-dimensional_1982, jain_composite-fermion_1989}, the $\nu=5/2$ state~\cite{willett_observation_1987}, the $0.7$ anomaly~\cite{thomas_possible_1996, kristensen_bias_2000, cronenwett_low-temperature_2002} or the Kondo effect~\cite{goldhaber-gordon_kondo_1998, kouwenhoven_revival_2001}, which are subject of current fundamental research. By confining interaction-induced states via quantum point contacts (QPCs)~\cite{miller_fractional_2007, dolev_observation_2008}, interferometers or quantum dots (QDs)~\cite{zhang_distinct_2009, ofek_role_2010}, one hopes to utilize experimental techniques like coherent charge- and spin-manipulation~\cite{hayashi_coherent_2003, petta_coherent_2005}, full counting statistics~\cite{gustavsson_counting_2006} or controllable coupling to other two-level systems~\cite{shinkai_correlated_2009, frey_dipole_2012} to gain further insight into the underlying physics. Many of these experiments rely on detecting changes of a localized charge state via an adjacent QPC serving as the charge detector~\cite{field_measurements_1993, elzerman_few-electron_2003, vandersypen_real-time_2004}. Its detection fidelity poses a fundamental limit to the readout speed of qubits and has therefore been subject of several investigations. It was found that the detection fidelity can be improved by maximizing both the detector's pinch-off slope and the capacitance between QD and QPC~\cite{zhang_engineering_2004}. The detector slope was increased by employing SETs~\cite{lu_real-time_2003, ilani_microscopic_2004, martin_localization_2004} or QDs~\cite{barthel_fast_2010} as detectors. The capacitive coupling can be improved by placing QD and detector in close proximity to each other~\cite{shorubalko_self-aligned_2008, choi_correlated_2009}, by avoiding metal gates in-between QD and detector~\cite{rossler_highly_2010} and/or by employing a floating gate on top of QD and detector.

Unfortunately, efficient charge detection is difficult to accomplish in $\rm{Al}_x \rm{Ga}_{1-x}\rm{As}$-based ultra-high-mobility two-dimensional electron systems (2DESs) due to the methods employed for achieving the high mobilities.
\begin{enumerate}
\item Screening layers between 2DES and doping layer were found to suppress remote ionized impurity scattering~\cite{friedland_new_1996, umansky_extremely_1997} but it comes at the cost of  hysteresis effects and temporal drifts when investigating electrostatically defined structures~\cite{miller_fractional_2007, dolev_observation_2008, rossler_gating_2010}. Moreover, transport experiments on QPCs defined in high-mobility 2DESs suggest increased screening of the confinement potential due to the presence of screening layers~\cite{rossler_transport_2011} which might further reduce the capacitive coupling between QD and detector.
\item High-mobility 2DESs are usually defined deeper underneath the surface ($\gtrsim200\,\rm{nm}$) compared to standard 2DESs ($\lesssim100\,\rm{nm}$). The increased distance to metal gates reduces screening, but being farther away from the surface reduces the coupling between QD and detector due to the higher dielectric constant of GaAs compared to vacuum. Moreover, since the typical distance between QD and detector scales with the distance between 2DES and surface gates, the coupling is expected to decrease further.
\item Many experiments on high-mobility 2DESs aim at confining (fractional) quantum Hall states. If the sample is investigated in the quantum Hall regime, the Hall resistance is added to the detector circuit's impedance. Hence, a given change of the detector's transmission results in a smaller absolute change of the detector current.
\end{enumerate}

\section{Charge Detection Experiments}
\begin{figure}
\includegraphics[scale=1]{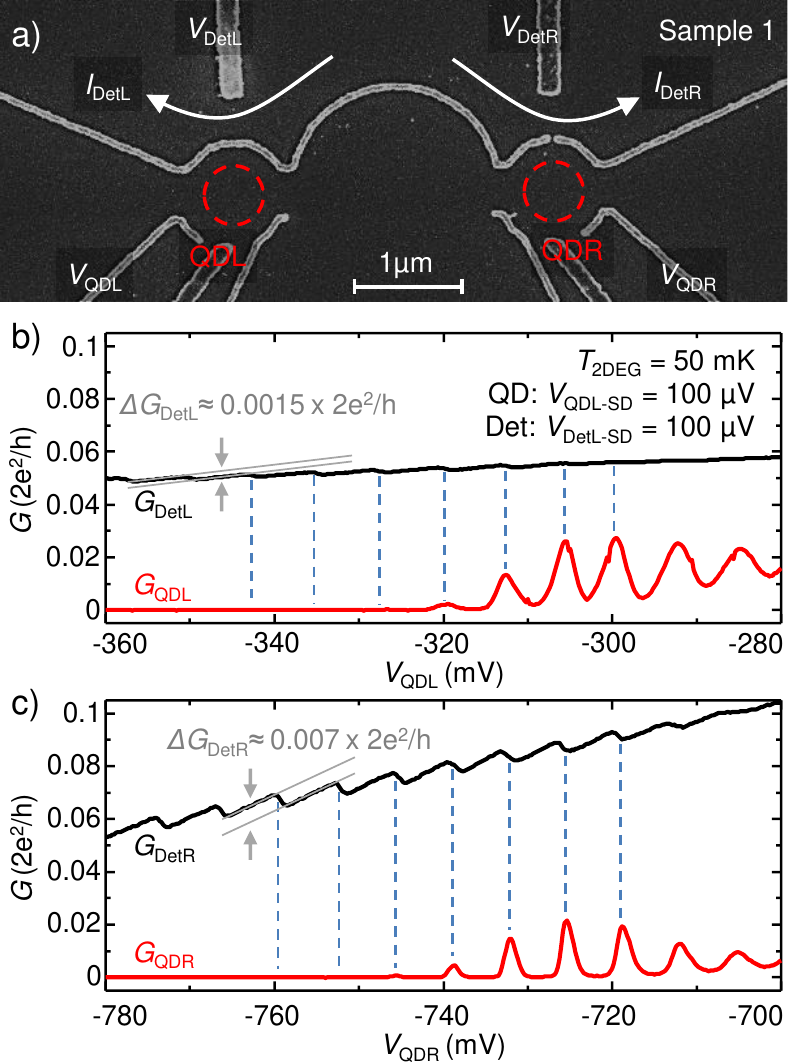}
\caption{\label{fig:detection}(color online) a) Scanning electron micrograph of Sample 1. Schottky gates appear bright, the crystal surface appears dark. Applying negative voltages to gates depletes the underlying electron gas and defines two QDs (dashed circles) and two constrictions serving as charge detectors (white arrows). The left and right half of the structure are identical with the exception of a gap in the gates between the right QD and its detector. b) Gate voltage $V_{\rm{QDL}}$ controls the number of electrons in the left QD, giving rise to Coulomb blockade oscillations in the differential conductance $G_{\rm{QDL}}=dI_{\rm{QDL}}/dV_{\rm{QDL-SD}}$ (red). Each change of the electron occupation number is accompanied by a kink in the differential conductance of the detector $G_{\rm{DetL}}=dI_{\rm{DetL}}/dV_{\rm{DetL-SD}}$ by typically $\Delta G_{\rm{DetL}}\sim0.0015\times\rm{2e^2/h}$ (black trace). c) Differential conductance of the right QD $G_{\rm{QDR}}$ and the right detector $G_{\rm{DetR}}$, plotted as a function of gate voltage $V_{\rm{QDR}}$. The high detector sensitivity of $\Delta G_{\rm{DetR}}\sim0.007\times\rm{2e^2/h}$ is attributed to the particular geometry of the detector gates, as will be shown later.}
\end{figure}

We present a study of several charge detectors realized in high-mobility 2DESs. The 2DES is defined at a $\rm{GaAs}/\rm{Al}_{\it{x}} \rm{Ga}_{\it{1-x}}\rm{As}$ heterointerface, $z=120\,\rm{nm}$ (Samples $1,2,3$) or $z=320\,\rm{nm}$ (Sample $4$) beneath the surface. The doping is realized in a region with reduced aluminium content $x=0.24$, providing high mobilities without using additional screening layers~\cite{gamez_$nu5/2$_2011}. Typical electron sheet densities and mobilities without LED-illumination are ($n_{\rm{S}}=1.4\times10^{11}\,\rm{cm^{-2}}$, $\mu=9\times10^{6}\,\rm{cm^{2}V^{-1}s^{-1}}$, Samples $1,2,3$) and ($n_{\rm{S}}=1.3\times10^{11}\,\rm{cm^{-2}}$, $\mu=10\times10^{6}\,\rm{cm^{2}V^{-1}s^{-1}}$, Sample $4$). After performing optical lithography to define mesa, Ohmic contacts and gate leads, the inner part of the structure is defined by electron beam lithography and subsequent deposition of Ti/Au gates as shown in the scanning electron micrograph of Sample $1$ in figure~\ref{fig:detection}a).
After cooling the sample to a temperature of $T_{\rm{el}}\approx 50\,\rm{mK}$, applying negative voltages to the gates (bright areas) depletes the 2DES underneath and defines two QDs and two detectors. In order to compare the same geometry under different screening conditions, the barrier gate separating the right detector from the right QD is split which should reduce screening between QD and detector. Figure~\ref{fig:detection}b) shows the differential conductance $G_{\rm{QDL}}=dI_{\rm{QDL}}/dV_{\rm{QDL-SD}}$ of the left QD (lower trace) as a function of the voltage applied to the QD plunger gate. Characteristic Coulomb blockade oscillations indicate the formation of a QD. The differential conductance of the detector is measured simultaneously and displays a step of $\Delta G_{\rm{DetL}}\sim0.0015\times\rm{2e^2/h}$ every time the charge on the QD changes by one electron. The magnitude of the step is a measure for the readout fidelity and was optimized by measuring the step height at different detector conductances. It was found that the readout step height is approximately constant for $0.01\times\rm{2e^2/h}\lesssim \it{G}_{\rm{Det}}\lesssim\rm{0.5}\times\rm{2e^2/h}$. A likely explanation for this observation would be the compensation between two effects: the steeper detector-slope at $G=0.5\times\rm{2e^2/h}$ versus reduced screening by the 2DES at lower conductance due to a reduced electron density in the vicinity of the detector. Since a lower detector current also results in less back-action from the measurement circuit on the QD~\cite{gustavsson_frequency-selective_2007,taubert_telegraph_2008,gasser_statistical_2009,kung_noise-induced_2009}, the detectors are set to a conductance of $G_{\rm{Det}}\sim0.1\times\rm{2e^2/h}$. The experiment is repeated with the right QD and a gapped detector and yields again Coulomb blockade oscillations (lower trace) and detector kinks (upper trace). This time the detector sensitivity is $\Delta G_{\rm{DetR}}\sim0.007\times\rm{2e^2/h}$. Comparing our detector sensitivities to the best reported values for top-gate defined single QDs $\Delta G\sim0.004...0.006\times\rm{2e^2/h}$~\cite{field_measurements_1993, vandersypen_real-time_2004, vink_cryogenic_2007, muller_radio-frequency_2011}, the left detector's performance appears to be below average, whereas the right detector is rather sensitive. There are several possible reasons for the increased sensitivity of the right detector. The presence of the gap should reduce both the electrostatic screening and the lateral distance between QD and detector. However, since the size of the gap ($\sim40\,\rm{nm}$) is smaller than the depth of the 2DES ($120\,\rm{nm}$), the relative change of the amount of screening metal and the reduced distance is unlikely to explain an increase of the sensitivity by a factor of four. It turns out that the gap can be used to define a localized state, which will be investigated in more detail later.

\subsection{Increased coupling via floating gate}
\begin{figure}
\includegraphics[scale=1]{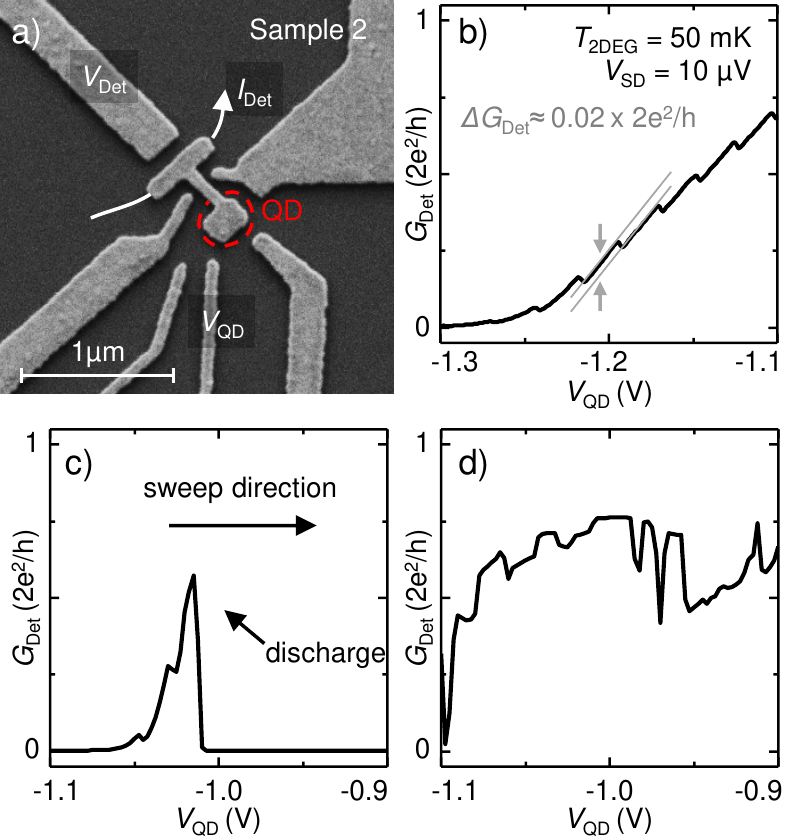}
\caption{\label{fig:floatinggate}(color online) a) Scanning electron micrograph of Sample 2. A floating gate on top of both the QD and the detector is employed to increase capacitive coupling between the two. b) Detector conductance as a function of gate-voltage $V_{\rm{QD}}$. Regular steps with $\Delta G_{\rm{Det}}\sim0.02\times\rm{2e^2/h}$ indicate charging of the QD measured by the detector. c) While repeatedly sweeping $V_{\rm{QD}}$, a sudden change of the detector's conductance is observed (black arrow). The implied presence of additional negative charge in the vicinity of QD and detector is probably caused by charging of the floating gate. d) After the discharge, the conductance of the detector fluctuates while sweeping $V_{\rm{QD}}$.}
\end{figure}

Further improvement of the detector sensitivity is expected by increasing the capacitive coupling between QD and detector via a floating gate. In the past, floating gates have been used between adjacent QDs~\cite{chan_capacitively_2003, hubel_two_2007, churchill_electronnuclear_2009} and were found to increase the capacitive coupling between them. Figure~\ref{fig:floatinggate}a) shows an image of Sample 2 which consists of QD, detector and a floating gate on top of them.
A typical charge readout signal is shown in figure~\ref{fig:floatinggate}b) as a function of the QD's plunger gate voltage. The detector sensitivity is now $\Delta G_{\rm{Det}}\sim0.02\times\rm{2e^2/h}$, roughly a factor of three better than literature values. However, in Sample 2 and two other similar samples (data not shown), severe charge rearrangements were observed within the first few days of the experiment. One example is shown in figure~\ref{fig:floatinggate}c). While measuring $G_{\rm{Det}}$ as a function of plunger gate voltage, a sudden decrease of $G_{\rm{Det}}$ is observed, corresponding to additional negative charge in the vicinity of the detector. Since the Schottky barrier between floating gate and GaAs should prevent vertical tunnel currents, the most likely explanation for such behavior is charging of the floating gate due to electrons tunneling laterally from a neighboring gate. After charging, $G_{\rm{Det}}$ is fluctuating at a timescale of seconds, as can be seen from the gate-sweep shown in figure~\ref{fig:floatinggate}d). Random switching events indicate time- and/or voltage-dependent charge rearrangements between the floating gate and its environment. These results indicate that despite improving the charge readout sensitivity, our floating gates are difficult to handle because the necessarily strong lateral electric fields induce charging events.

\subsection{Inducing a localized state in the detector}
\begin{figure}
\includegraphics[scale=1]{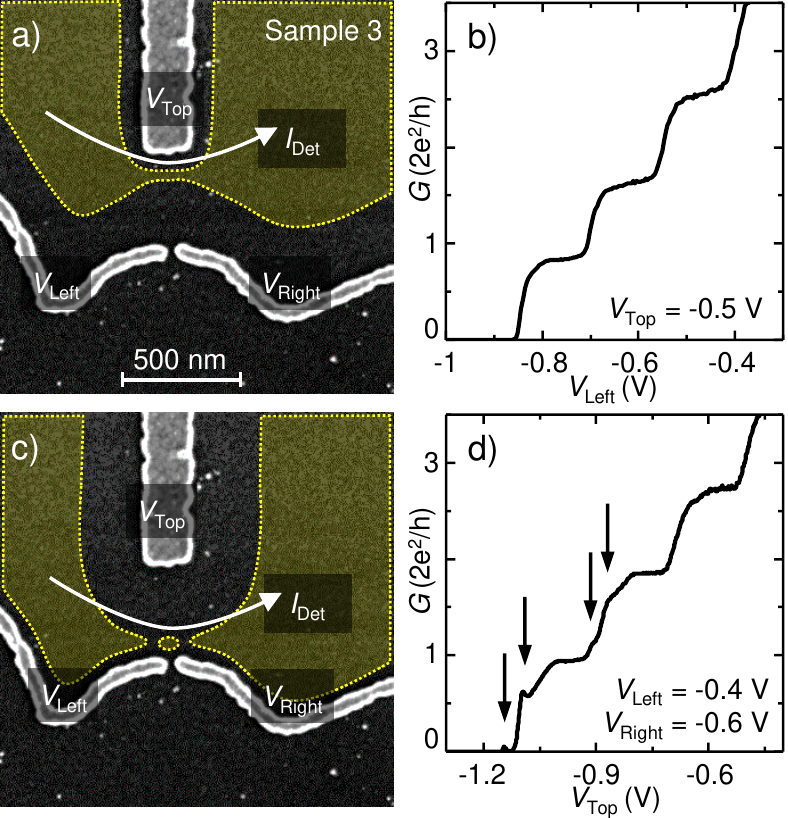}
\caption{\label{fig:lokalisierungscuts}(color online) a) Scanning electron micrograph of Sample 3. Applying more negative voltages to the bottom gates shifts the conducting channel (sketched yellow area) upwards. b) Linear conductance of the channel, plotted as a function of $V_{\rm{Left}}$ while $V_{\rm{Right}}=V_{\rm{Left}}-0.2\,\rm{V}$. After subtracting a serial resistance of $R_{\rm{S}}=10\,\rm{k\Omega}$ to account for cables, contacts and 2DES leads, the expected conductance quantization is observed. c) Applying more negative voltage to the top gate shifts the channel downwards, enabling a localization to form at the gap between the left and right gate. d) Linear conductance as a function of $V_{\rm{Top}}$. Multiple charging events of a localized state are observed (arrows).}
\end{figure}

\begin{figure}
\includegraphics[scale=1]{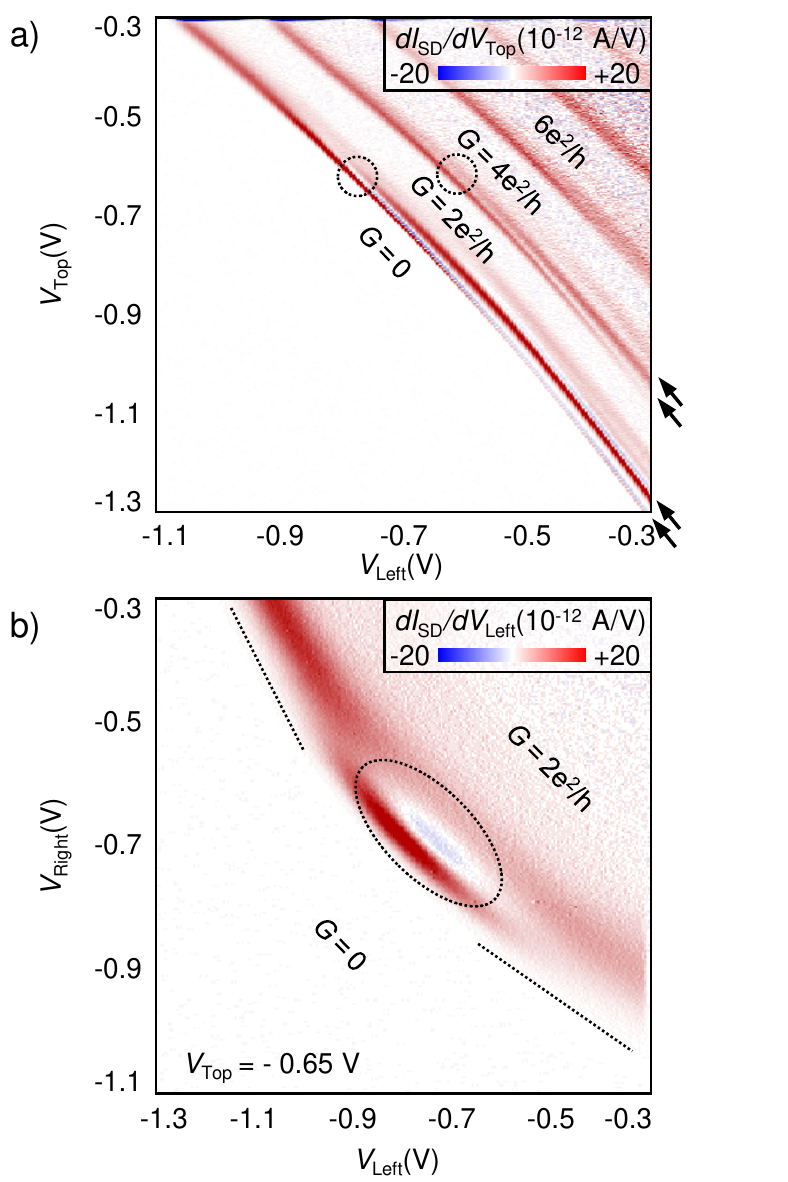}
\caption{\label{fig:lokalisierungsverschiebung}(color online) a) Transconductance $G_{\rm{T}}=dI/dV_{\rm{Top}}$ plotted in color scale as a function of $V_{\rm{Top}}$ and $V_{\rm{Left}}$ with $V_{\rm{Right}}=V_{\rm{Left}}-0.2\,\rm{V}$. The conductance within white areas is constant and reaches from zero (bottom left) to conductance values of $G=1,2,3\times \rm{2e^2/h}$. Additional dark lines (indicated by arrows) emerge when the channel is shifted downwards, closer to the bottom gates. Dashed circles mark the positions where the localization starts to be formed. The resonances' positions relative to the conductance plateaus remain unchanged, implying the localized state shifts together with the channel (rather than being fixed in space). b) Changing the ratio of the left and right tunnel barriers at constant $V_{\rm{Top}}=-0.65\,\rm{V}$. Two dashed lines indicate where the channel is pinched off between the right and top gate (bottom-right of graph) and between left and top gate (top-left of graph). When both tunnel barriers are defined, the localized state emerges (dashed ellipse).}
\end{figure}

Since the capacitive coupling can not easily be increased beyond using thin gates and introducing a gap in the barrier, we now focus on increasing the slope of the detector's pinch-off curve. Charge detection experiments on double QDs were previously performed using the Coulomb resonance of a third QD as detector~\cite{barthel_fast_2010}. Its slope depends on tunnel coupling and temperature and is much larger than values achievable with a typical saddle-point shaped constriction~\cite{buttiker_quantized_1990}. Shaping the electrostatic confinement potential of the detector is performed on Sample 3, which is lithographically identical to the right QD of Sample 1. Figure~\ref{fig:lokalisierungscuts}a) shows one top gate and two bottom gates which are used to tune the detector's confinement potential.
Keeping $V_{\rm{Top}}$ at a moderate voltage and sweeping the voltage applied to the bottom gates should result in an single constriction as sketched by the shaded region (yellow). The linear conductance of the constriction is plotted in figure~\ref{fig:lokalisierungscuts}b) as a function of the voltage applied to the bottom gates and displays the expected quantization in multiples of $\rm{2e^2/h}$. Figure~\ref{fig:lokalisierungscuts}c) illustrates the idea of pushing the detector closer to the gap between left and right gate. If the geometry and voltages are suitably chosen, it is possible to create a localized state. Figure~\ref{fig:lokalisierungscuts}d) shows the detector conductance as a function of $V_{\rm{Top}}$ while $V_{\rm{Left}}$ and $V_{\rm{Right}}$ are held constant. In contrast to the smooth conductance quantization observed earlier, the pinch-off curve exhibits several resonances (marked by arrows) indicating charging events of the localized state. In order to investigate the transition of the detector from QPC-like to QD-like, the full parameter sweep is shown in figure~\ref{fig:lokalisierungsverschiebung}a). The transconductance $G_{\rm{T}}=dI/dV_{\rm{Top}}$ is plotted in color scale (areas of constant conductance $G=0,1,2,3\times \rm{2e^2/h}$ appear white) as a function of $V_{\rm{Left\&Right}}$ and $V_{\rm{Top}}$.
Varying the ratio of $V_{\rm{Left\&Right}}$ and $V_{\rm{Top}}$ shifts the detector up ($V_{\rm{Top}}=-0.3\,\rm{V}$) or down ($V_{\rm{Top}}=-1.3\,\rm{V}$) as verified via scanning gate microscopy experiments on another sample~\cite{schnez_imaging_2011}. While the pinch-off curve is smooth when the detector is defined close to the top-gate, additional resonances appear in the middle of the plot (dashed circles) and become more pronounced as the detector is shifted towards the bottom gates. The resonances are parallel to each other, indicating that they have the same capacitance with respect to the gates and must therefore be multiple charging events of a single localized state. Moreover, the resonances are linked to the quantized plateaus. The localization's resonances are not clearly separated from the QPC's pinch-off curve which is compatible with the geometry of the gates: since the lateral distance between the left and right gate is only $\sim40\,\rm{nm}$, the potential minimum in-between all three gates should be rather small and shallow. The localized state can therefore not contain multiple well-separated charging events without being strongly coupled to its leads. It is also noteworthy that the $0.7$ anomaly is observed throughout the plot. It starts as a smeared area of finite slope around $G\sim0.7\times \rm{2e^2/h}$ (top left) and turns into a pronounced plateau with conductance $G\sim0.6\times \rm{2e^2/h}$ when the localized state is formed. Finite-bias data (not shown) also indicate the coexistence of the charged localized state with the characteristic half-plateaus of the $0.7$ anomaly. For future studies, it might prove interesting to investigate this peculiar transition from an open QPC to a localized state as it might shed light on the microscopic nature of the $0.7$ anomaly~\cite{cronenwett_low-temperature_2002, meir_kondo_2002, reilly_phenomenological_2005}. However, this topic is not further pursued in this paper.

In addition to moving the localization up/down, as a further test it can be moved left/right by varying the ratio of $V_{\rm{Left}}$ and $V_{\rm{Right}}$ at constant $V_{\rm{Top}}$. The resulting plot of the transconductance $G_{\rm{T}}=dI/dV_{\rm{Right}}$ is shown in figure~\ref{fig:lokalisierungsverschiebung}b). Two dashed lines mark the pinch off between left and top gate (top-left of graph) and between right and top gate (bottom-right of graph). Only when both tunnel barriers are equally close to pinch-off, the localized state is formed (dashed ellipse). The resonances' slope of approximately $45^\circ$ means that the localized state has about the same capacitance and hence distance to the left and right gate which agrees with the schematic drawing shown in figure~\ref{fig:lokalisierungscuts}c).

The observation of a single set of resonances implies that the magnitude of the sample's disorder potential is comparable or weaker than the variations of the gate-induced confinement potential. If the disorder potential would dominate over the variations of the gate-induced confinement potential, one would expect resonances with random spatial position and hence random slopes when shifting the detector channel up/down or left/right. In contrast, our resonances stem from the estimated location of the gate-induced potential minimum with the additional requirement that the localization's tunnel barriers must be defined via gates. This observation indicates a different origin of the localized state than defect-induced resonances observed in samples with lower mobility~\cite{mceuen_resonant_1990}.  Still, in our high-mobility devices the gate-induced potential operates on top of a (weak) disorder-induced potential background. As a speculative microscopic scenario, the gate-induced potential might modify a local potential minimum by tuning it into a more localized state. Either way, the presented geometry can be used to tune a QPC-like detector into a localized state that can be employed as a very sensitive charge detector. Defect-induced localized states were used to increase the sensitivity in other charge detection experiments, for example in graphene- and indium-arsenide-based devices~\cite{guttinger_charge_2008, shorubalko_self-aligned_2008}. In contrast, our method applied with the low defect density of high-mobility heterostructures provides control of the position and coupling of the localized state.
\subsection{Tuned charge detector in the quantum Hall regime}
\begin{figure}
\includegraphics[scale=1]{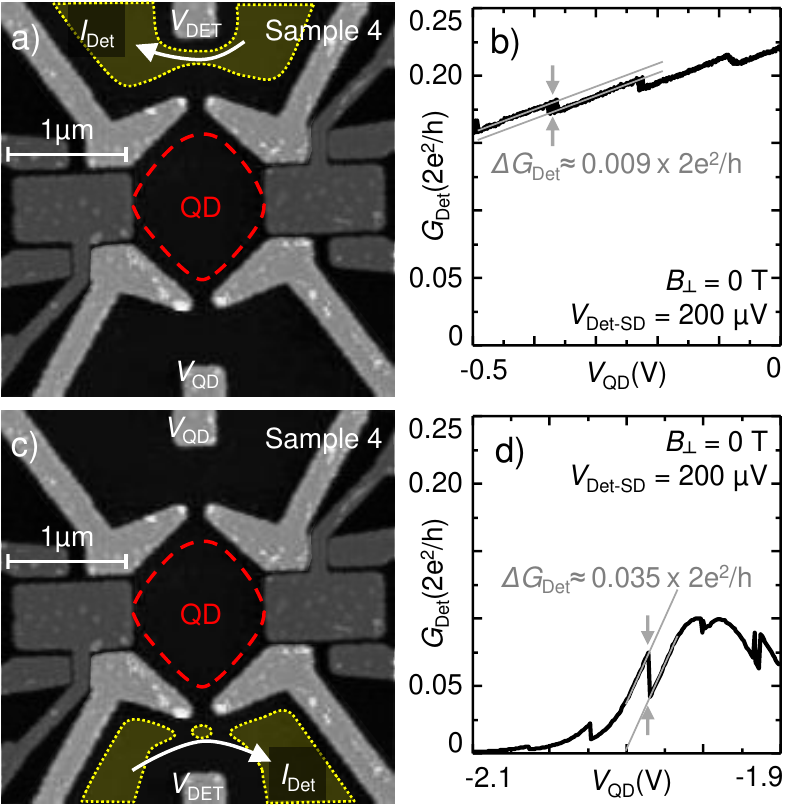}
\caption{\label{fig:qpc-vs-localization}(color online) a) Atomic force micrograph of Sample 4. Four gates define a QD (red). The detector channel (yellow) forms a QPC which is tuned to the steepest part of its pinch-off curve. b) Differential conductance of the QPC as a function of $V_{\rm{QD}}$. Charge detection steps of $\Delta G_{\rm{Det}}\sim0.009\times\rm{2e^2/h}$ indicate a well-tuned detector. c) The bottom charge detector is tuned into a localized state (yellow island). d) The differential conductance of the charge detector exhibits a resonance with superimposed charge detection steps. At the position with the largest slope, charge detection steps are separated by $\Delta G_{\rm{Det}}\sim0.035\times\rm{2e^2/h}$, demonstrating an improvement by roughly a factor of four over the non-localized charge detector.}
\end{figure}

\begin{figure}
\includegraphics[scale=1]{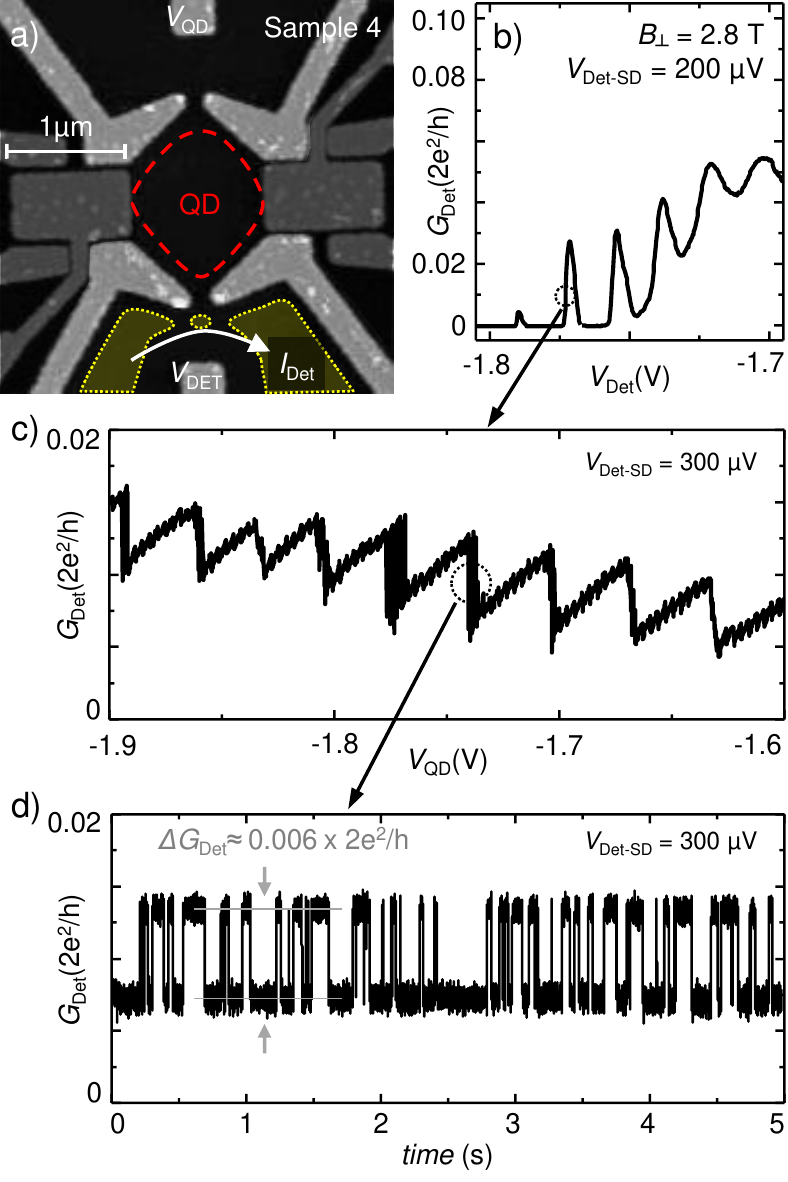}
\caption{\label{fig:quantumhalldetection}(color online) a) Atomic force micrograph of Sample 4. Four gates define a QD (red), the detector channel (yellow) is tuned into the localized regime. b) Differential conductance of the detector-localization as a function of $V_{\rm{Det}}$. The position marked by a dashed circle is used for charge detection. c) $G_{\rm{Det}}$, plotted as a function of $V_{\rm{QD}}$. Big steps correspond to a change of the QD occupation by one electron. Small wiggles are caused by the simultaneous sweep of $V_{\rm{Det}}$ in order to keep the detector on the steep side of the localization's resonance. At the transition marked by a dashed circle, $G_{\rm{Det}}$ is measured with $1\,\rm{kHz}$ bandwidth as a function of time, as shown in d): A telegraph signal is created by electrons tunneling through the QD. The occupation numbers are separated by $\Delta G_{\rm{Det}}\sim0.006\times\rm{2e^2/h}$.}
\end{figure}

The technique of forming a localized state in a QPC was found to work in different geometries and heterostructures. Figure~\ref{fig:qpc-vs-localization}a) shows an AFM-image of Sample 4, where experimental results had to be obtained under relatively difficult experimental conditions: a large QD, rather far away from the detector and defined in the quantum Hall regime. In order to compare different detector settings, the QD is first characterized without applying a magnetic field.
Five gates define QD (red circle) and detector (yellow area), another distant gate is used as QD plunger gate. Since the 2DES of Sample 4 is defined $320\,\rm{nm}$ beneath the surface, the lateral distances are scaled up compared to the previous samples. The gate voltages are chosen such that the detector's pinch-off curve does not exhibit resonances. Charging events of the QD are visible as steps of the detector's differential conductance shown in figure~\ref{fig:qpc-vs-localization}b). After careful optimization, the highest obtainable readout fidelity was $G_{\rm{Det}}\sim0.009\times\rm{2e^2/h}$ which is comparable to other well-tuned QPC charge detectors~\cite{field_measurements_1993, vandersypen_real-time_2004, vink_cryogenic_2007, muller_radio-frequency_2011}. The same sample configured such that the bottom detector employs a localized state is schematically shown in figure~\ref{fig:qpc-vs-localization}c). The detector's pinch-off curve displays multiple resonances with the sharpest one shown in figure~\ref{fig:qpc-vs-localization}d). Charge events of the QD appear as steps superimposed on the resonance. At the steepest point of the resonance, the readout step height reaches $G_{\rm{Det}}\sim0.035\times\rm{2e^2/h}$ which exceeds the charge detection fidelity achieved with our floating gate. Since the geometry of the sample is symmetric, these observations imply that with comparable screening conditions and mutual capacitances, localized states strongly improve the readout fidelity. By employing the localized state as a detector, it is possible to perform charge detection experiments with high accuracy even in the quantum Hall regime, as shown in figure~\ref{fig:quantumhalldetection}a).
A magnetic field of $B_\perp=2.8\,\rm{T}$ is applied perpendicular to the 2DES, corresponding to quantum Hall filling factor $\nu=2$. After tuning the detector into the localized regime, Coulomb blockade oscillations of the detector are observed as a function of $V_{\rm{Det}}$, as shown in figure~\ref{fig:quantumhalldetection}b). The steepest slope (dashed circle) is used for charge detection of the QD. Figure~\ref{fig:quantumhalldetection}c) shows the detector conductance as a function of $V_{\rm{QD}}$, while $V_{\rm{Det}}$ is swept as a compensation to keep the detector at the flank of its Coulomb resonance. Periodic small wiggles are due to the finite resolution of the compensating voltage source, whereas large steps are caused by charging events of the QD. The tunnel rates of the QD are slow enough to observe them in real time, giving rise to telegraph noise at the border between adjacent occupation numbers. At one of these borders (dashed circle), the detector current is recorded with $1\,\rm{kHz}$ bandwidth as a function of time and plotted in figure~\ref{fig:quantumhalldetection}d). Clear two-level behavior separated by $\Delta G_{\rm{Det}}\sim0.006\times\rm{2e^2/h}$ is observed, demonstrating the high sensitivity of localized charge detectors. As a comparison, measuring the charge state with a QPC-like detector (data not shown) we observe conductance level separations of $\Delta G_{\rm{Det}}\sim0.002\times\rm{2e^2/h}$. In future studies, we hope to extend the charge detection experiments to the regime of fractional quantum Hall states and to investigate time-dependent processes of such systems.

\section{Conclusion}
We investigated several methods of improving the sensitivity of charge detectors. The capacitive coupling between QD and detector was increased by using a floating gate. However, the increased sensitivity comes at the cost of charge rearrangements, making this technique difficult to handle in a typical gate-defined nanostructure. Introducing a gap in the barrier gates between QD and detector, we find a strongly enhanced sensitivity which is attributed to reduced screening, reduced lateral distance between QD and detector and a steeper detector slope due to the formation of a localized state. Formation and lateral shifting of the localized state was investigated and demonstrates that the detector can be tuned gradually from QPC-like to QD-like characteristics. Finally, the technique of using a localization for sensitive charge-readout was applied to a large QD in the quantum Hall regime.

\section{Acknowledgments}
We acknowledge the support of the ETH FIRST
laboratory and financial support of the Swiss Science Foundation
(Schweizerischer Nationalfonds, NCCR Quantum Science and Technology).
\section*{References}
\bibliographystyle{iopart-num}
\bibliography{Bibliothek}

\end{document}